\documentclass[12pt,twocolumns]{iopart}

\usepackage{iopams}
\usepackage{epsfig}
\begin{document}

\title{A null test of the metric nature of the cosmic acceleration}
\author{A. Buzzi$^1$, C. Marinoni$^1$ and S. Colafrancesco$^{2,3}$}

\address{$^1$Centre de Physique  Th\'eorique, UMR 6207 CNRS-Luminy and Universit\'e de Provence,
 Case 907, F-13288 Marseille Cedex 9, France France\\
 $^2$ ASI-ASDC, c/o ESRIN, Via G.Galilei, I-00040 Frascati, Italy\\
 $^3$ INAF - Osservatorio Astronomico di Roma, via Frascati 33, I-00040 Monteporzio, Italy}

\ead{buzzi@cpt.univ-mrs.fr, christian.marinoni@cpt.univ-mrs.fr, cola@mporzio.inaf.it}

\begin{abstract}

We discuss the testable predictions of a phenomenological model in which the
accelerated expansion of the universe is the result of the action of a
non-gravitational force field, rather than the effect of a negative-pressure
dark-energy fluid or a modification of general relativity.  We show, through the equivalence 
principle, that in such a scenario the cosmic acceleration felt by distant standard candles like
SNIa (type Ia Supernovae (SNe)) depends  on the mass of the host
system, being larger in isolated galaxies than in rich clusters. As a consequence, the
scatter in the observed SNIa Hubble diagram has mostly a physical origin in
this scenario: in fact, the SNIa distance modulus is increasing, at fixed
redshift, for SNe that are hosted in isolated galaxies with respect to the case
of SNe hosted in rich galaxy clusters. Due to its strong dependence on the
astrophysical environments of standard candles, we conclude that alternative
non-gravitational mechanisms for the observed accelerated expansion of the
universe can be interestingly contrasted against the standard metric
interpretation of the cosmological acceleration by means of an environmental
analysis of the cosmic structures in which SNIa are found. The possible absence
of such environmental effects would definitely exclude non-gravitational
mechanisms being responsible for the accelerated cosmological expansion and will
therefore reinforce a metric interpretation.
\end{abstract}


\maketitle

\section{Introduction}
\label{intro}

The unprecedent convergence of observational results that we are currently
witnessing has narrowed down the region of the cosmological parameter space
which is compatible with all the different and independent probes of cosmology:
Supernovae \cite{Riess98} \cite{Perlmutter99}, CMB \cite{deb00, Spergel07} and Large
Scale Structures \cite{Tegmark06, Guzzo08, Mar08}. Robustly growing evidence suggests that {\it
i)} ordinary matter is a minority ($\sim 1/6$)  of all the matter content of
the universe, {\it ii)} matter -- mostly dark, non-baryonic
matter -- is a minority ($\sim 1/4$) of all the cosmological mass--energy
density, mostly contributed by an obscure form of energy referred to as `dark
energy', {\it iii)} the 3D spatial geometry of the universe is flat and {\it
iv)} the expansion of the cosmic metric has been accelerating for the last $\sim 7$
Gyrs of our universe lifetime.

Even though the big picture is in place, the two dominant
contributions to the stress-energy tensor -- i.e. dark energy and dark matter
-- still have a hypothetical nature and they have  not been discovered
yet.
While there is widespread evidence for the existence of the non-baryonic dark matter
component producing the potential wells of large-scale structures \cite{Clowe06},
as yet no persuasive theoretical explanation has been able to elucidate the
physical nature of the dark energy component \cite{pebrat}.
As a matter of fact, unveiling the nature of dark energy and its role in cosmology and
gravitation is a difficult and subtle challenge.
In such a context, it should not be overlooked that the large roaming from model to model,
and the abundance of weakly predictive theories, might eventually limit the possibility
of discriminating between different competitors proposed so far for explaining the observed
dynamics of the accelerating universe.

In the absence of a compelling theoretical explanation for the dark energy
component, and in a minimal, zero-order approach, we explore here  the
possibility of preserving the standard metric interpretation of the accelerated
expansion against possible alternative physical scenarios.
To this end, we first evaluate and then discuss the observable consequences of
local, non-gravitational mechanisms which could in principle  accelerate matter
in our Hubble patch of the universe. We assume here that the universe is described by
general relativity, that it is dominated by components which satisfy the usual
energy conditions (according to which the universe can only decelerate) and that
the onset of recent accelerated expansion  is the result of the presence of a
hypothetical non-gravitational force field.
Such an alternative explanation is rather conservative, since it assumes
neither a cosmological constant (or negative-pressure fluid) nor a modification
of general relativity.
Accordingly, we first work out a self-consistent, non-geometric model for the
cosmic acceleration that is able to reproduce the current observations of
standard candles (i.e. SNIa) and then we discuss a falsifiability procedure
aimed at testing its observational predictions.

The motivation behind this work is to put strong limits on a hypothetical (or
non usually considered) physics that is possibly missing in our picture of the
universe, and, in turn, to strengthen the evidence supporting the standard
paradigm with which we are currently explaining its past history, its present
stage and its future fate.

\section{Accelerated cosmological expansion with a non-cosmological,
large-scale, radial force field}
Our goal is to work out testable predictions that allow us to reject
non-metric acceleration models. To this end, we construct here a general,
phenomenological model in which the role of dark energy is mimicked by an
alternative mechanism of non-gravitational origin.

We consider here a background universe with a metric expansion as predicted by
general relativity. We assume  that in such a universe a hypothetical
large--scale, non-gravitational force field influences the overall dynamics of
large scale structures in a patch of the universe with typical dimensions of
the local Hubble volume. In this scenario every object which at time $t$ sits
on the shell of a sphere of proper radius $r(t)$ centered on the observer feels
a peculiar acceleration field $\gamma_p(t)$ that is radially directed and time
dependent.\\
We further speculate that the only cosmological component contributing to the
stress-energy tensor is dark matter. In other words, we assume that there is no
dark energy at all in the universe and that what we interpret as apparent
isotropic acceleration of the metric is indeed the combination of the decelerated cosmological expansion predicted by general relativity plus the Doppler effect sourced by matter  which is
accelerating outward under the effect of the radial force field we have added
to this cosmological scenario.

In such a case,  by appropriately tuning the non-cosmological contribution to
the observed redshift, one can reconstruct a functional form  for the
luminosity distance $d_L$ (see eq. \ref{eq.dlgeneral} below) which reproduces
the standard one derived within a model with cosmological constant (or dark
energy): i.e.
\begin{equation}
d_{L}(z_{obs}, {\bf p}_{obs})=d_{L}(z,{\bf p}, {\bf \gamma_p}) \; .
 \label{dleq}
\end{equation}
In other words, the standard $\Lambda$CDM cosmological parameter set (represented by the
vector ${\bf p_{obs}}$) observationally inferred by simply plugging in the observed
redshift $z_{obs}$ into the luminosity distance formula are biased with respect to the
{\it true cosmological} values (represented by the vector ${\bf p}$) that would be
naturally inferred by recognizing the presence of a physical mechanism responsible for the
acceleration of matter {\it in situ}.

Having outlined the phenomenological model we are pursuing, let us now
elucidate its finer details, i.e. the dependence of the luminosity distance
(Eq. 1) on the peculiar acceleration term $\gamma_p$.
\\*
The luminosity distance of an object standing at the observed redshift
$z_{obs}$ is given, in a pure matter scenario, by
\begin{equation}
 d_L = (1+z_{obs})  {c \over H_0 \sqrt{|\Omega_k|}} S_k
 \bigg[ \sqrt{|\Omega_k|} \int_0^{z_{obs}} \frac{dz} {E(z)}  \bigg] \; ,
 \label{eq.dlgeneral}
\end{equation}
where $\Omega_k= 1- \Omega_m$, $S_k(x) = \sin(x)$ (if $k=1$), $S_k(x)= x$ (if
$k=0$), $S_k(x)= \sinh(x)$ (if $k=-1$), and $E(z)=[\Omega_m (1+z)^3 + \Omega_k
(1+z)^2]^{1/2}$.\\
While in the standard model  the measured redshift $z_{obs}$ has a pure
cosmological interpretation, in the accelerated model it additionally includes
the contribution of the peculiar velocity $v_p$ induced by the radial non
gravitational field and it can be written as
\begin{equation}
z_{obs}=z+\frac{v_p}{c}(1+z).
\label{eq.zo}
\end{equation}
Clearly, there are an infinite variety of cosmological models that, when
combined with an equally arbitrary variety of radial acceleration fields
$\gamma_p$, could in principle reproduce the observational features of the
standard cosmological model. So, for the sake of simplicity, and in order to
capture the essential physics of the problem, we assume in the following that
the universe is flat, that its total density is contributed only by matter
$\Omega_0=\Omega_m=1$, and its expansion rate is
$H(z)=H_0(1+z)^{\frac{3}{2}}$.
\\*
Under these hypotheses, the luminosity distance can be written as
\begin{equation}
d_{L}(z, v_p)=\frac{2c}{H_0} \Big\{ \Big(1+z)\Big(1+\frac{v_p}{c}\Big)
\Big\} \Big\{ 1-\Big[(1+z)\Big(1+\frac{v_p}{c}\Big)\Big]^{-\frac{1}{2}}\Big\}
\; ,\label{eq.dl_om1}
\end{equation}
\noindent nonetheless, we stress that these working assumptions do not influence the generality
of the conclusions presented in $\S 3$: analogous considerations hold, in fact,
for a low density universe, open universe (with, e.g., $\Omega_m \sim 0.2$).

We can now compute how a peculiar velocity $v_p$ in this scenario
depends on the peculiar acceleration field: $v_p=v_p[\gamma_p(t)]$ at
a given cosmological redshift $z$.
\\*
Using the expression of the proper distance to a given coordinate,
$r(t)=a(t)x(t)$, we obtain the expression of the acceleration $\gamma(t)$ by
means of which a perturbed metric expands radially:
 \begin{equation}
\gamma=\frac{\ddot{a}}{a}r+\frac{\dot{a}}{a}v_p+\frac{dv_p}{dt} \; ,
 \label{eq.alpha}
 \end{equation}
where $v_p=a\dot{x}$ and the peculiar acceleration term is given by
\begin{equation}
\frac{dv_p}{dt}+Hv_p=\gamma_p(t) \; .
 \label{eq.dwdt}
\end{equation}
After transforming this equation from the time to the redshift domain, we find
the general solution
\begin{equation}
 v_p(z)=(1+z)\Big(K - H_0^{-1} \int \frac{\gamma_p(z)}{(1+z)^{7/2}}dz\Big) \; .
 \label{eq.wz}
\end{equation}
We model the peculiar acceleration term in this equation by means of a general, non-divergent
polynomial function
\begin{equation}
\gamma_p(z)=\gamma_0+\sum_{i=1}^n{\gamma_i(1+z)^{-i}}
 \label{eq.gamma}
\end{equation}
where the $\gamma_i$ are  free parameters to be adjusted with the constraint given
by Eq. 1. Note that in the absence of a peculiar acceleration acting on cosmic
matter, we recover the standard result that any primordial peculiar velocity is
damped by the background expansion (i.e., $v_p \propto a^{-1}$).

\subsection{Fitting the data}
We show here that by appropriately tuning the parameters $\gamma_i$  we can
match the luminosity distance $d_L$ of standard candles like SNIa as inferred
within the standard model of cosmology.
    \\*
To this end, we first set the integration constant $K=0$ by imposing that, in
the limit $z\rightarrow \infty$, the peculiar velocity  $v_p=0$. We also set
$\gamma_0=0$ in order to avoid any primordial acceleration field. Then, we fit
Eq. \ref{eq.dlgeneral} to reproduce the currently available supernovae data. We
find that even a flat, matter--dominated ($\Omega_m=1$) cosmological model, when
supplied with an appropriate local acceleration field, would be able to
reproduce the luminosity distance derived by fitting the observed data assuming
(`{\it wrongly'}) a cosmological origin of the source redshift (see
Fig.\ref{fig2}). \\

We obtain an acceptable fit ($\chi_{\nu}^2 \sim 1$) by expanding up to order $n=4$ in Eq. 8.
The best fitting coefficients  in the redshift
range $0<z<1.7$ (for a background cosmological model with $\Omega_m=1$) are:
$\gamma_1=2.1(\pm 0.5)\cdot
10^{-8}$m/s$^2$, $\gamma_2=8.7(\pm 0.6)\cdot 10^{-8}$m/s$^2$, $\gamma_3=1.1(\pm 0.5)\cdot
10^{-7}$m/s$^2$, $\gamma_4=-4.3(\pm 0.3)\cdot 10^{-8}$m/s$^2$.

\begin{center}
\begin{figure}[h]
\includegraphics[width=75mm,angle=0]{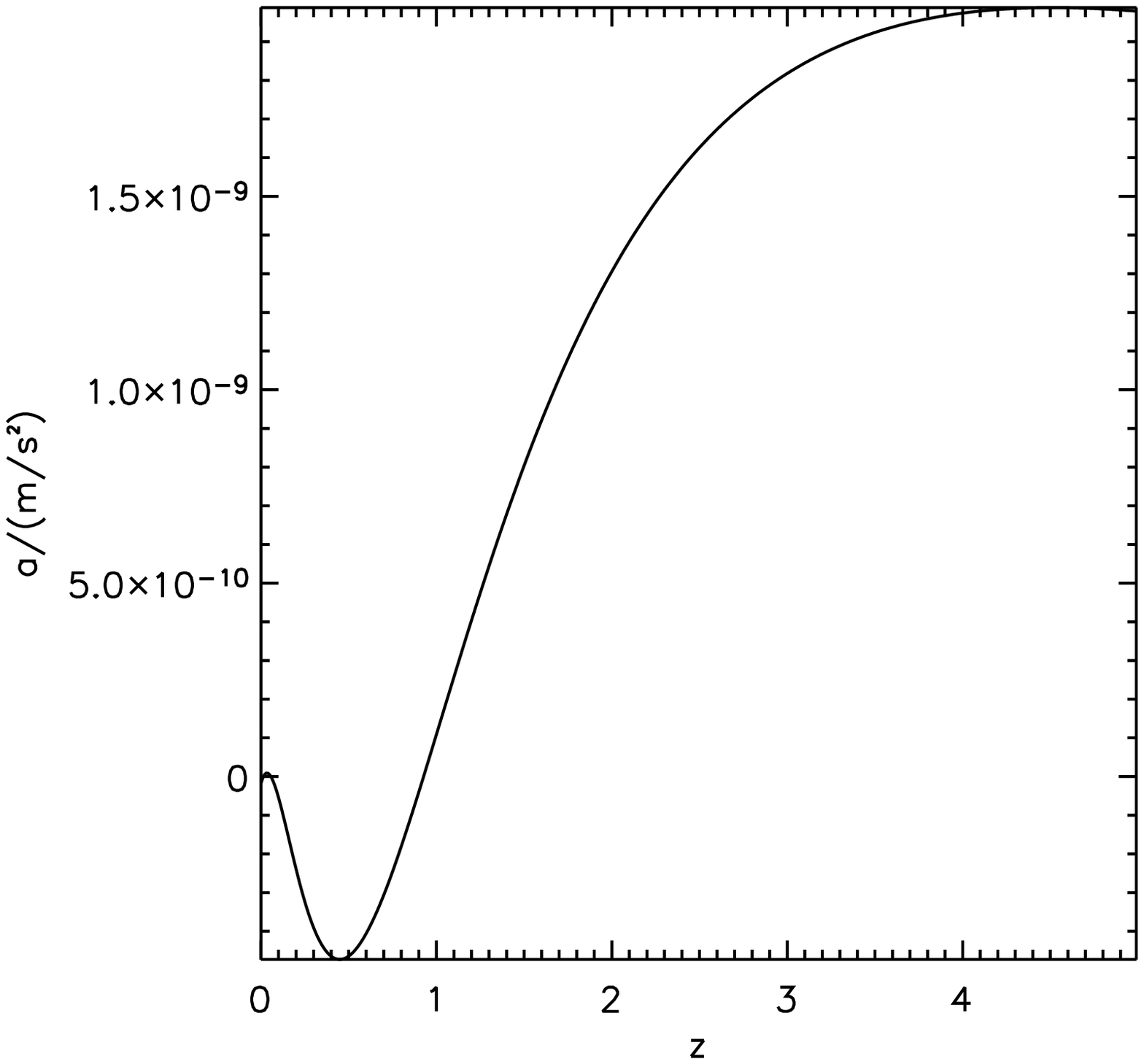}
\includegraphics[width=68mm,angle=0]{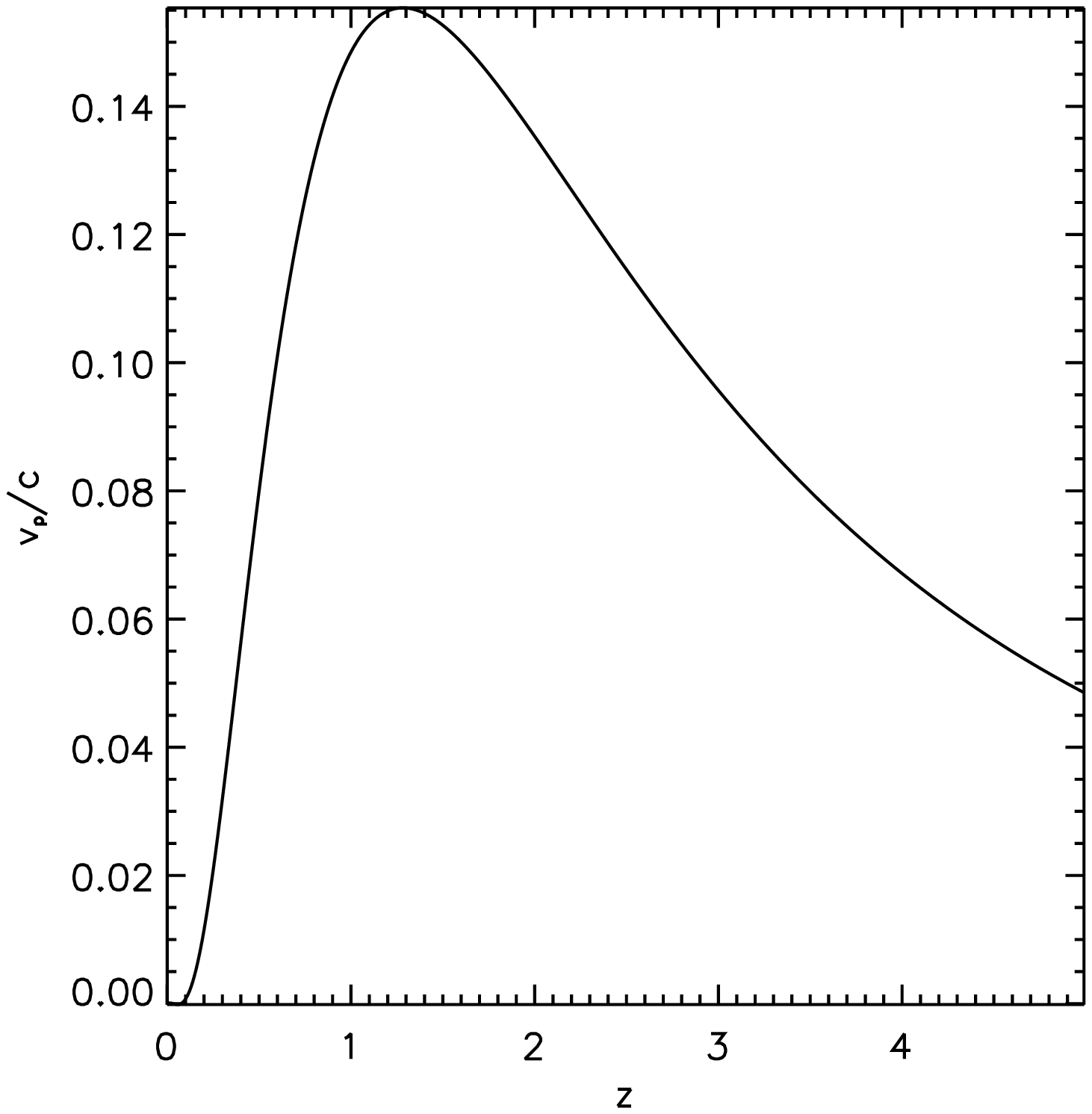}
\caption{The best fitting peculiar acceleration $\gamma_p$ {\it (left)} and
peculiar velocity $v_p$ {\it (right)} needed to reproduce current SNIa data (in
the redshift range $0<z<1.7$) are shown as a function of redshift in a flat,
decelerating $\Omega_m=1$ background cosmology.} \label{fig 1}
\end{figure}
\end{center}

\begin{center}
\begin{figure}[h]
\includegraphics[width=75mm,angle=0]{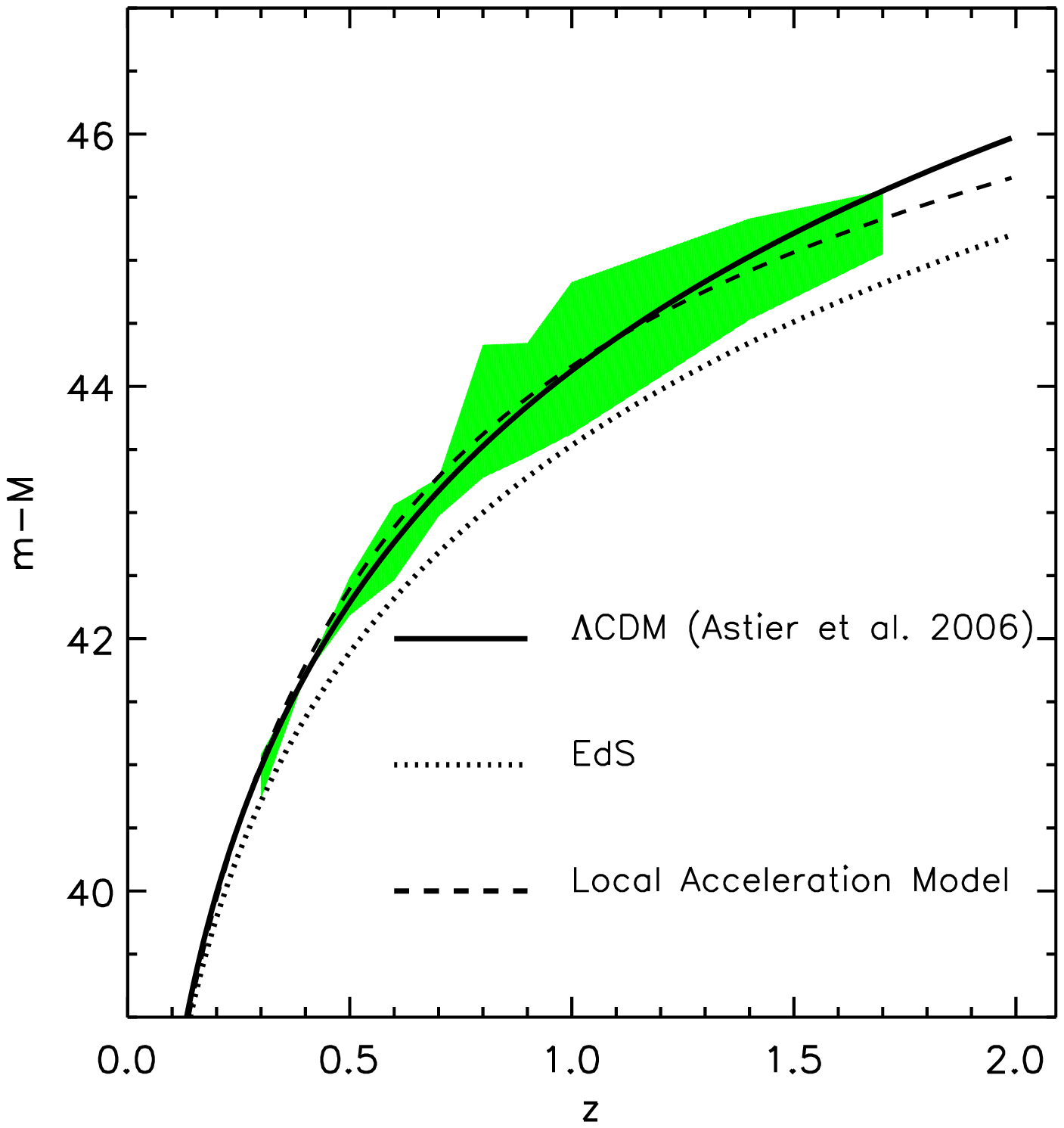}
\includegraphics[width=75mm,angle=0]{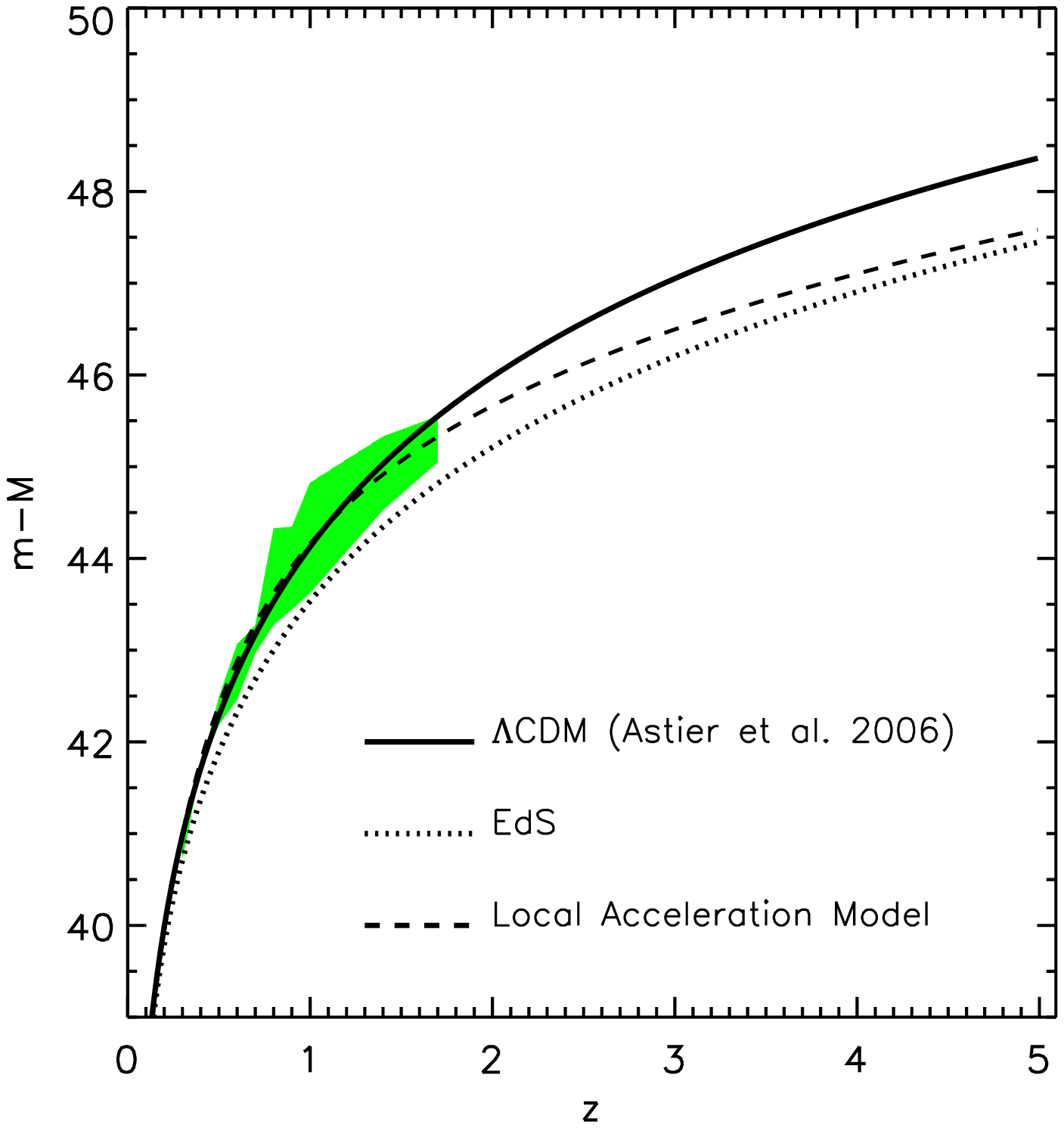}
\caption{{\em Left :} the SNIa distance modulus is shown as a function of redshift for
different cosmic expansion models: the best fitting  $\Lambda$CDM model
obtained by using distant supernovae  \cite{Riess07} (solid), a decelerating
Einstein deSitter model (dotted) and a model in which a large scale peculiar
acceleration field acts on top of the EdS world model (short-dashed curves).
The shaded area shows the state-of-the-art dispersion in the observed SNIa
samples from the SNLS collaboration \cite{Astier06}  and from the HST gold sample
\cite{Riess07}. {\em Right :} We show the convergence, at high redshift,  of the peculiar
acceleration model to the underlying background cosmological model (an Einstein de Sitter
model in this case).} \label{fig2}
\end{figure}
\end{center}

We find that the large scale peculiar acceleration required to fit the SNIa
data (see Fig.\ref{fig2}) is negligible with respect to other local
acceleration fields. Given the smallness of its amplitude, it modifies almost
undetectably the inertial nature of the reference systems to which it is
applied at high redshift.
Note that, in general, this is the order of magnitude of any large-scale, peculiar acceleration field one needs to introduce in order to reinterpret SNIa data over the redshift range $0 < z < 1.7$.
For $z \to \infty$, the distance modulus predicted  within this paradigm converges to the standard
behavior expected in the underlying, decelerated, cosmological model because the peculiar perturbation vanishes at early epochs.

We stress here that, by definition, the acceleration induced by such an
hypothetical non-gravitational force should act differentially on the baryonic
and non-baryonic matter components. Therefore, any possible non-gravitational
accelerating  mechanisms are physically viable only if it does not destroy
(e.g., by segregating it into parts) the astronomical object on which they act
(for instance, by segregating the dark matter halo from the luminous baryonic
component hosted within it). In other words, the acceleration time scale
$\tau_{acc}$ in a cosmological frame has to be much larger than the typical
dynamical time scale $\tau_{dyn}$ of a system trapped inside a dark matter halo.\\
By inserting the time-averaged  value  of Eq. 8
($\langle \gamma_p \rangle \approx 5.7 \cdot 10^{-10} m s^{-2}$, if
$\Omega_m=1$) into Eq. 5 and solving by using $v_p=a\dot{x}$, we obtain hence
\begin{equation}
\tau_{acc} \equiv  \bigg[2.9 {R H_0^{2/3} \over  \langle \gamma \rangle }
\bigg]^{3/4} \approx (G\rho)^{-1/2} \; ,
\end{equation}
where $R$ is the comoving radius of the dark matter halo assumed in virial
equilibrium, and $\rho$ is its mean density. This simple physical argument puts
stringent constraints on the viability of such a class of models. Unless one is
willing to consider non standard model of gravity sourced only by baryons
[e.g., MOND (Milgrom 1983) or TeVeS theories (Bekenstein 2004)], or as yet
unexplored interactions mediated by the dark matter particles, one can use the
previous argument to doubt the physical reliability of such
non-gravitational acceleration mechanism.
Moreover, models of non-cosmological acceleration that are able to reproduce the
metric acceleration  without invoking the effect of dark energy are not
positively defined over all the epoch of cosmological interest. Specifically,
to explain the acceleration of distant standard candles without violating local
constraints, one would need to invoke an acceleration of opposite sign at low
and high redshift (see the left panel of Fig. 1). Additionally, these models yield
also quite high peculiar velocities, which can even be of the order of a tenth
of the speed of light at a characteristic epoch of $z \sim 1-2$ (see the right
panel of Fig. 1). In principle, the strong sensitivity of the value of the
large-scale peculiar velocity in such scenarios can also be used to put
constraints on the viability of the non-gravitational acceleration models.

On one hand, one must mention the ability of these models to handle the
age-problem in FRW cosmologies with no dark energy. A recurring argument for a
dark energy component has come from considerations that the observed stellar
ages are too long to be consistent with the observed Hubble parameter
$H_0\sim 70$ km s$^{-1}$ Mpc$^{-1}$ for an universe which has always been decelerating. Since
our results have been presented, for the sake of illustration, for a flat CDM
model, they are consistent with the stellar estimates of the age of the
universe only if low values of $H_0\sim 50$ km s$^{-1}$ Mpc$^{-1}$ are viable \cite{Blanc03}.

Although there are authors claiming -- given the still large systematic
uncertainties affecting $H_0$ measurements --  that at present the age argument
cannot be used to definitively exclude the Einstein de Sitter model and,
therefore, arguing for a cosmological constant \cite{Row02} \cite{Sark08}, we
note here that the non gravitational acceleration models can naturally
reconcile decelerated background cosmological models  with the age of the
oldest stars in the universe.

As a matter of fact the observed Hubble constant evaluated by comparing the observed
redshift ($z_{obs}$) with  redshift-independent distance indicators (d) would
overestimate the true universal value $H_0$ obtained by using, as prescribed by
theory, the cosmological redshift (z).
The SNIa Hubble diagram is constructed in such a way to be independent of the 
values of $H_0$. Anyway, one can directly estimate the amplitude of the $H_0$ bias.  
In the accelerated model the expansion rate in the local 
universe  is overestimated by a factor  $\delta H_0=v_p/d$.
Therefore, at $z\sim 10^{-2}$,  
the isotropic peculiar expansion needed to recover 
a Hubble parameter of $\sim 50$ km s$^{-1}$ Mpc$^{-1}$ is of order of $v_p/c=10^{-2}$. 

We stress here the analogy between this '{\it reconciliation mechanism}' and
the one invoked within the `Hubble bubble' scenarios. In fact, in such models
\cite{Zehavi} we are supposed to live inside an underdense region (Hubble
bubble) that is expanding faster than the average. Measurements of the Hubble
constant within the underdense region would therefore overestimate the
universal value by $\delta \rho/\rho=-3\delta H/H$ thus offering a different
way to bypass the age problem within a decelerating background metric.

We will see in the next session, however, that a crucial observational test
can be devised to reject such non-gravitational  acceleration mechanisms.

\section{Testable predictions}

We now  discuss the observable effects that a hypothetical
non-gravitational force field, say a `{\it dark force field'}, should have  on
the global expansion of the universe, as traced by a set of distant standard
candles (e.g., SNIa). We show that the physical imprints of such a
{\it `dark force'} can be unambiguously contrasted  with those of a {\it
`dark energy'} component. In particular, it is possible to  discriminate
between them directly in the Hubble diagram rather than, as usually done, in
the equation of state parameter space.
Indeed, if such a non-gravitational, radial force field exists (see, e.g., the
preliminary suggestion that a large-scale magnetic tension might mimic the
effects of dark energy \cite{Contopoulos07}, or the equivalent proposal of a
large local void \cite{Calsteb}), we can show that it accelerates cosmological
objects (test particles) at the same radial distance in a different way. In
fact, since only the gravitational force has the property to depend on the
gravitational mass -- which via the equivalence principle is equal to the
inertial mass -- it follows that, in the proposed `{\it dark force}' scenario
SNe are not inertial systems in free-fall along geodesics of the space-time.
The acceleration of SNe will thus have a unique signature: it will depend on
the mass of the hosting system. Since in this {\it dark force} scenario the
luminosity distance $d_{L}$  of a cosmic object depends on its acceleration
(see eq. \ref{eq.dl_om1} and \ref{eq.wz} ), the different inertial masses of
the test particles will cause predictable deviations from the mean, best
fitting curve shown in the left panel of Fig. 2.

One can parameterize the mass dependence of the peculiar acceleration experienced
by a given system $\gamma_p^*$ in terms of the best fitting mean acceleration
$\bar{\gamma}_p$ using a simple scaling law
\begin{center}
\begin{equation}
\gamma_p^*=\Bigg(\frac{\bar{M}}{M^*}\Bigg)^{\alpha} \bar{\gamma}_p
\label{acmo}
\end{equation}
\end{center}
\noindent where
the exponent $\alpha$ characterizes the specific physical mechanism responsible for
the acceleration of supernovae. The limiting case $\alpha=0$ describes the action
of the gravitational field showing  that, in this case,
the acceleration of test particles is mass-independent.
In contrast, the acceleration generated by non-gravitational force fields is generically
described by setting  $\alpha \neq 0$. In particular, the case $\alpha=1$ illustrates the
large class of force fields in which the acceleration is inversely proportional
to the mass of the test particle (for example models in which  the strength of the
force is independent of the  mass of the object experiencing the field).

\begin{figure}[h]
\begin{center}
\includegraphics[width=140mm,angle=0]{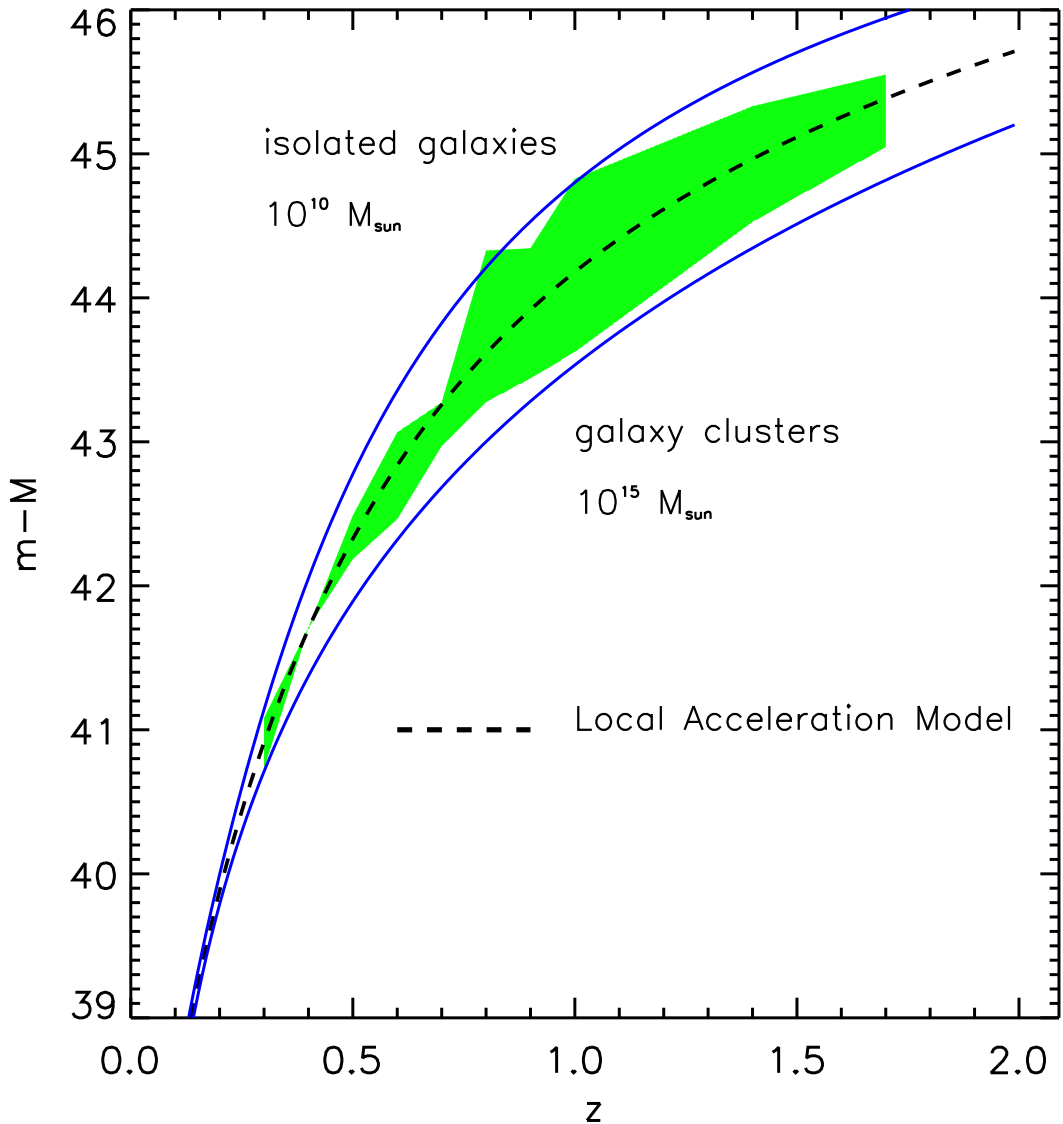}
\caption{The SNIa distance modulus is shown as a function of redshift in the
`{\it dark force}' scenario (black dashed curve). The lower (upper) solid
curves bracketing the extremal envelopes of the dispersion in the Hubble
diagram (shaded green area) are obtained by  assuming $\alpha=1$ in Eq.
\ref{acmo} and by letting the minimum and maximum mass of the test particles
(SNe host systems) vary over five orders of magnitude i.e., from normal isolated
galaxies to galaxy clusters.}
 \label{fig3}
 \end{center}
\end{figure}

For the sake of simplicity let us discuss first  the case $\alpha=1$. A direct
consequence of this assumption is that, at fixed redshift, SNe hosted in isolated
normal galaxies (with typical mass of $\sim 10^{10} h^{-1}M_{\odot}$), when subject to
the force field we are challenging here, will experience an acceleration of
order $\sim 10^{5}$ stronger than SNe hosted in rich galaxy clusters (with
typical mass of $\sim 10^{15} h^{-1}M_{\odot}$). (We assume here that the
masses of the host systems remain constant with time). This effect
systematically shifts the distance modulus of SNe  in low (high) mass hosts
towards higher (lower) values with respect to the reference, best fitting curve
characterizing supernovae hosted in some fiducial system of mass $\bar{M}$ (see
Fig. \ref{fig3}).

Similarly, and without loss of generality, we find that for all the
values $|\alpha| \gtrsim 0.1$, (i.e. even in the extreme cases of an
acceleration field which has a weak dependence on the host mass), the family of
distance moduli associated to SNe hosted in systems whose mass differs by up to  
five orders of magnitude still spans the whole dispersion region characterizing
current data. The systematic deviation  from the best fitting relation of the
distance moduli of supernovae hosted in a large variety of  systems, going from
small to large masses,  is therefore practically insensitive to the exponent
$\alpha$ parameterizing the mass dependence of the acceleration felt by
supernovae. This general result follows  from the fact that a) the distance
modulus calculated by including the peculiar acceleration term cannot  be
smaller than  the one computed in the associated background cosmological model,
and b) to each value of the exponent $\alpha$ one can always associate an
appropriately tuned value of the fiducial mass in the range
$10^{10}<\bar{M}/M_{\odot}<10^{15}$  in such a way that  a change of five orders of
magnitude in the mass of the host gives distance moduli which are always
confined between the upper and lower envelopes of the dispersion in the SNIa
Hubble diagram.

Therefore, we  conclude that in a {\it dark force} scenario, part of the
scatter in the SNIa Hubble diagram has a physical origin: for a fixed redshift
$z$ and $\alpha \gtrsim 0.1$ ($\alpha \lesssim -0.1$), SNIa with larger values
of $m-M$ in Fig. \ref{fig3} should be hosted in small (big) systems, while SNIa
with smaller values of $m-M$ should be hosted in rich clusters (isolated normal
galaxies).

To summarize, a null-test of the metric nature of the accelerated expansion can
be easily performed by means of an environmental analysis of the cosmic
structures in which SNIa are found. Such a study could be optimally performed
by future large-area sky surveys of distant SNe such as SNAP \cite{alde06}. The
strong sensitivity to the host system mass of the test strategy that we suggest here
will allow to exclude any hypothetical non-metric large-scale interaction which
could be in principle responsible for the observed kinematics of the universe.
Such a result would hence shed a much clearer light on the nature of the
assumed dark energy component dominating the late stages of the cosmic
evolution.

 \vskip 1.truecm

 \noindent {\bf Acknowldgements}. We thank the referee for useful comments.
We acknowledge stimulating and useful
discussions with D. Fouchez, P. Taxil and J.M. Virey. S.C. thanks
the Centre de Physique Th\'eorique de Marseille for hospitality.


\end{document}